\documentclass[12pt]{iopart}
\usepackage{graphicx}

\expandafter\let\csname equation*\endcsname\relax
\expandafter\let\csname endequation*\endcsname\relax
\usepackage{amsmath,amssymb}  
\usepackage{bm}  
\usepackage{cite} 
\usepackage[titletoc,toc,title]{appendix}
\usepackage{graphicx}
\usepackage{subfig}
\usepackage{caption}
\usepackage{sidecap}
\usepackage{simplewick}

\begin{document}
\title[Learning of couplings for random asymmetric kinetic Ising models revisited]
{Learning of couplings for random asymmetric kinetic Ising models revisited: random correlation matrices and
learning curves}
\author{Ludovica Bachschmid-Romano
and Manfred Opper}
\address{Department of Artificial Intelligence, Technische Universit\"{a}t Berlin,
Marchstra{\ss}e 23, Berlin 10587, Germany}
\ead{ludovica.bachschmidromano@tu-berlin.de and manfred.opper@tu-berlin.de}

\begin{abstract}
We study analytically the performance of a recently proposed algorithm 
for learning the couplings of a random asymmetric kinetic Ising model 
from finite length trajectories of the spin dynamics. Our analysis
shows the importance of the nontrivial equal time
correlations between spins induced by the dynamics for the speed of learning.
These correlations become more important as the spin's stochasticity is decreased. 
We also analyse the deviation of the estimation error 
from asymptotic optimality.
\end{abstract}


\section{Introduction}
Recently, the learning of synaptic couplings for a recurrent neural network 
modelled by a kinetic Ising model with random couplings has attracted attention in the statistical physics community, 
see e.g \cite{Mezard_Sakellariou_2011, Roudi_Hertz_prl,Roudi_JStat, Aurell_2012_PhysRevE, Kabashima_2013, Dunn_2013,
Tyrcha_Hertz_2014, Bachschmid-Romano_Opper_2014,  Saad_2014, Battistin_2015}.
The model is defined by
a system of N Ising spins $ \sigma_i$ connected through couplings $J_{ij}$. 
We assume throughout the paper that the interactions are non--symmetric, i.e. we have $J_{ij} \neq J_{ji}$
and $J_{ii} = 0$.
The system evolves in discrete time according to a synchronous parallel dynamics, where spins at time $t+1$
are updated independently with transition probability (specialised on the case of no external fields)
\begin{equation}
P(\sigma_i(t)|\{\sigma_j(t-1)\}_{j=1}^N)  = \frac{e^{\beta \sigma_i(t) \sum_j J_{ij} \sigma_j(t-1)}}
{2 \cosh (\beta \sum_j J_{ij} \sigma_j(t-1) )}.
\label{eq:dynamics}
\end{equation}
We are interested in learning the spin couplings $J_{ij}$, assuming that a complete trajectory $\{ \bm{ \sigma} \}_{0:T}= \{{\sigma}_i(t)\}_{i=1,\ldots, N, t=1,\ldots,T}$ of length $T$ for all spins is observed.  A well known solution to this problem is given by the method of maximum likelihood, which leads to a set
of coupled nonlinear equations which have to be solved by iteration. A computationally much simpler and elegant solution 
valid for large networks with random couplings which avoids an iterative solution was recently presented in \cite{Mezard_Sakellariou_2011}. 
This solution is based on an exact mean field (EMF) expression for spin correlations  which can be explicitly solved for the 
couplings. The EMF estimator replaces exact correlations by empirical correlations which can e.g. be computed from 
a single spin trajectory. Simulations have shown good agreement between true and estimated couplings \cite{Mezard_Sakellariou_2011}.

Of course, if there is only a limited number of observations available there will be a nonzero estimation error
for the EMF method. One may then ask  how much one has to pay for the numerical efficiency of the algorithm
in terms of a loss in statistical efficiency. Hence, we would like to investigate
at what rate the error decreases 
with growing length of trajectories and if the decrease is slower than that of a statistically  efficient
estimator such as the maximum likelihood estimator which has an optimal asymptotic rate \cite{Schervish_book}.
Using the replica method we will compute the estimation error of the EMF method in the thermodynamic limit
$N\to\infty$ assuming that the data are generated from a kinetic Ising model with true couplings drawn at random from
a Gaussian distribution. The analysis of the statistical properties is significantly simplified
by the fact that kinetic Ising models with non--symmetric random couplings have spin correlations which decay
after a single time step (see for example \cite{Eissfeller-Opper}) and computations of learning curves resemble those for temporally independent data.
A nontrivial aspect however is the occurrence of equal time spin correlations of the spin dynamics. 
We compute an exact result for the statistics of the random correlation matrix. From this it is possible to obtain
 an explicit expression for the learning curve for the EMF algorithm and the asymptotics of the ML estimator. 

\section{Estimators}
\label{sec:estimators}
The EMF estimator \cite{Mezard_Sakellariou_2011}  is based on a linear relation between the time-delayed and the equal time correlator matrices,
\begin{equation}
C_{ij} = \langle \delta\sigma_i(t) \delta\sigma_j(t) \rangle, \qquad
D_{ij} =  \langle \delta\sigma_i(t+1) \delta\sigma_j(t) \rangle,
\end{equation}
 for the spin fluctuations $\delta\sigma_j(t) \doteq \sigma_j(t) - m_j(t)$, where
$m_j(t)$ denotes the local magnetisation at time $t$ and the brackets $\langle \dots \rangle$ denote expectation with respect to the spin dynamics (\ref{eq:dynamics}). Here we assume stationarity for which the matrices are time independent. 
If the couplings $J_{ij}$ are assumed to be mutually independent Gaussian random variables, with zero meand and variance $1/N$, the following mean field relation is found to be
 exact in the thermodynamic limit $N\to\infty$ :
\begin{equation}
D_{ij} = a_i  \sum_k J_{ik} C_{kj},
\label{MF_pred}
\end{equation}
where 
\begin{equation}
a_i =  \beta \int {\cal{D}}x \left[1- \tanh^2[\beta(H^{\text ext} + x \sqrt{\Delta_i})]\right],
\qquad \Delta_i=\sum_j J_{ij}^2 (1-m_j^2)
\end{equation}
and
${\cal{D}}x$ is the normal Gaussian measure. 
Throughout the paper we will specialise to the case of 
zero external field and vanishing initial magnetisations. In this case
we have $m_i(t) = 0$, $H^{\text ext} = 0$, $\Delta_i = 1$ and $a_i =a$ is independent of time.  
For the estimator the exact correlation matrices $\bm{C}$ and $\bm{D}$ 
are approximated by empirical averages 
using a long trajectory of spins (assuming zero magnetisations):
\begin{equation}
C_{ij} \rightarrow \hat{C}_{ij}= \frac{1}{T}\sum_{t=1}^T \sigma_i(t) \sigma_j(t), \qquad
D_{ij} \rightarrow \hat{D}_{ij}=\frac{1}{T}\sum_{t=1}^T \sigma_i(t+1) \sigma_j(t).
\label{eq:empirical}
\end{equation}
One can then obtain the couplings by inverting 
(\ref{MF_pred}) as follows:
\begin{equation}
J_{ij} = \frac{1}{a} \sum_k \hat{D}_{ik} \hat{C}^{-1}_{kj}.
\label{eq:EMF_espl_est}
\end{equation}
It is easy to see that the EMF estimator 
can be rephrased as the minimiser of the following cost function
\begin{equation}
E^{i}_{MF}= \frac{1}{2}\sum_{t=1}^T \left(\sigma_i(t) - a \sum_j J_{ij} \sigma_j(t-1) \right)^2 
\label{eq:E_EMF}
\end{equation}
with respect to the couplings $\{J_{ij}\}_{j=1}^N$. 
Note that 
the estimation of the ingoing couplings $\{J_{ij}\}_{j=1}^N$ for each spin $i$ can be treated separately
for the coupling distribution we are considering.
The EMF estimator is based on simple explicit computation 
(inversion of the correlation matrix in \ref{eq:EMF_espl_est}, which is possible if the parameter $\alpha = T/N$ is grater than $1$)
which makes the method fast. Other
estimators such as the well known maximum likelihood method (ML) have to resort to numerical
optimisations using iterative algorithms which could become computationally involved for 
large system sizes $N$ and a large number of data $T$. The ML estimator maximises
the probability of spin histories $\{ \bm{ \sigma} \}_{0:T}$ given by
\begin{equation}
P(\{ \bm{ \sigma} \}_{0:T}| \bm{J}) = \prod_{i=1}^N \prod_{t=1}^T  P(\sigma_i(t)|\{\sigma_j(t-1)\}_{j=1}^N) \; P(\sigma(0)),
\end{equation}
where  $P(\sigma(0))$ is the initial probability of spins. Since this probability factorises in the spins $i$ and $J_{ij}$ are assumed independent, 
the ML estimator for all couplings $\{J_{ij}\}_{j=1}^N$ pointing into spin $i$ minimises the cost function
\begin{equation}
E^{i}_{ML} = \sum_{t=1}^T \left (- \beta \sigma_i(t) \sum_j J_{ij} \sigma_j(t-1) +
\ln {2 \cosh (\beta \sum_j J_{ij} \sigma_j(t-1) })\right).
\label{eq:E_ML}
\end{equation}
While minimizing the cost function (\ref{eq:E_EMF}) just requires the computation of 
the empirical averages $\hat{\bm{C}}$ and $\hat{\bm{D}}$,
in order to minimize (\ref{eq:E_ML}) with respect to $J_{ij}$ one needs to compute the quantity $\sum_t \sigma_j(t) \tanh(\beta \sum_j J_{ij} \sigma_j(t))$
that explicitely depends on the current value of $J_{ij}$ and has to be recomputed at each step of the algorithm,
adding a $N_{\text step} \cdot T$ operation to the calculation. 
We observe that in order to avoid second order methods in the solution we need a fine tuning of the 
step size which makes the algorithm fairly slow for large $N$.
Although it is more computationally expensive, the ML estimator has the important property that it is asymptotically (i.e. for $T\to\infty$) 
{\em efficient}. This means that the asymptotic 
convergence of the mean squared estimation error to zero (assuming the model is correct)  happens
at a rate which is minimal for any (asymptotically) unbiased estimator \cite{Schervish_book}. 
In the following we will compute the error of the EMF algorithm in the thermodynamic
limit $N,T\to\infty$, keeping $\alpha$ fixed and compare with the asymptotic  $\alpha\to\infty$ 
optimal error rate of the ML estimator.

\section{Learning curves from the replica approach}
\label{sec:learning}
In this section we will introduce the replica method for computing the EMF  prediction error as a function
of the scaled number of observed data. We will work in a teacher--student scenario \cite{Engel_book, Opper_Kinzel_1996},
where the data are assumed to be generated at random from the dynamics of a teacher network 
with random couplings $J_{ij}^*$. We will use the scaling  $J_{ij}^* = W_{ij}^*/\sqrt{N}$ and assume that the $W_{ij}^*$ are 
independent Gaussian random variables with
$
W_{ij}^* \sim {\cal{N}}(0,1).
\label{prior}
$
We can treat the estimation of the ingoing couplings $\bm{W}^* \equiv \{W_{ij}^*\}_{j=1}^N$ for each spin $i$ separately.
For the sake of simplicity, in the following we will drop the index $i$ and define $W_j \doteq W_{ij}$.
The average square prediction error for any estimator of the couplings given by $\bm{W}$ is defined as
\begin{equation}
\varepsilon = \frac{1}{N} \overline{|| \bm{W}^* -  \bm{W} ||^2} =
1-2 \rho + Q,
\label{eq:def_err}
\end{equation}

where we defined 
\begin{equation}
\rho= N^{-1} \overline{\bm{W}^* \cdot \bm{W}}, \qquad
Q= N^{-1} \overline{||\bm{W}||^2} .
\label{eq:def_rho}
\end{equation}
The bar denotes an average over the spin trajectories $\{ \bm{ \sigma} \}_{0:T}$ generated with couplings $\bm{W}^* $ and over the
teacher couplings.
We will now analyse the performance  of algorithms which minimise a cost function of the type 
\begin{equation}
E=\sum_{t=1}^T \mathcal{E}(\sigma(t), h_t),  \qquad h_t=\frac{1}{\sqrt{N}} \sum_j W_j \sigma_j(t-1),
\end{equation}
such as (\ref{eq:E_EMF}) and (\ref{eq:E_ML}), on a random finite set of spin trajectories of size $T$. 
One can compute average properties such as the order parameters $\rho$ and $Q$ by introducing
an auxiliary probability density of couplings,
\begin{align}
& q(\bm{W}) = \frac{1}{Z} \; e^{-\nu E(\bm{W})},
\label{auxiliary}
\end{align}
with a formal inverse 'temperature' parameter $\nu$ and the partition function
\begin{align}
& Z(\bm{ \sigma})= \int d\bm{W}  e^{-\nu E(\bm{W})}.
\end{align}
For any $\nu$, we can compute disorder averages of 'thermal averages' of
variables such as $\rho$ and $Q$ from the quenched average of the
free energy per coupling, defined by
\begin{equation}
F=- N^{-1}\nu^{-1} \overline{ \log Z(\bm{ \sigma}) } =
-\nu^{-1} \lim_{n \to 0} \frac{\partial}{\partial n} N^{-1} \log  \overline{ Z^n(\bm{ \sigma})}.
\label{eq:free_def}
\end{equation}
By taking finally the limit $\nu \to \infty$ (zero 'temperature'), the probability density (\ref{auxiliary})
concentrates at the minimum of $E(\bm{W})$ and we can extract the desired order parameters.
To compute the average, we will make the following assumptions. While the spins $\sigma_i(t)$  are still treated as binary
random variables,  in computing expectations over $\sigma_j(t)$ for $j\neq i$ we assume a central limit theorem to be valid 
for the fields  $h_t$ as sums of a large number of weakly dependent random variables. Hence, 
we consider only the second order statistics of these variables and treat them as Gaussian random variables.
For equal times the corresponding Gaussian density would be 
$
p(\{\sigma_j(t)\}_{j\neq i}) = {\cal{N}}(0,\bm{C})
$,
where the stationary covariance matrix $\bm{C}$ is a random matrix
which itself depends on the random matrix of teacher couplings $\bm{W}^*$ of the entire network. 
For different times $t\neq t'$, dependencies between spins $\sigma_j(t)$
and $\sigma_k(t')$ are neglected. This is in accordance with our previous assumptions for $|t- t' | > 1$, but
we need an extra argument to justify neglecting $D_{jk}$ giving the correlations at times $t$ and $t+1$.
In principle, $\bm{D}$ might enter the computation of order parameters as well. (\ref{MF_pred}) shows a relation between the $\bm{D}$ and $\bm{C}$ matrices involving
 the teacher couplings linearly. The arguments presented later in section \ref{sec:stat_C} indicate that for the asymptotic random matrix calculations
involving similar relations we can treat teacher couplings and random matrices $\bm{C}$ as asymptotically  independent. 
Hence, we argue that in an expectation over teacher couplings the contributions due to $\bm{D}$ vanish.
We will see later that the statistical properties of the matrix $\bm{C}$ will enter the final result of the learning curve through
the self averaging moment $C_{-1} \doteq \frac{1}{N} Tr \overline{\bm{C}^{-1}}$. We will then show in 
section \ref{sec:stat_C} how this and other moments can be computed.
Thus we will include the average over the teacher couplings $W_{kj}$ for $k\neq i$ in the statistics of $\bm{C}$, but we 
need to perform the average over the teacher couplings $W^*_j \equiv W_{i j}^*$ pointing to spin $i$ explicitly. 
Finally, the dependencies between random correlation matrices $\bm{C}$ at different times are also neglected for $N\to\infty$. 
This results in an effective  statistical weight over spin histories given by
\begin{equation}
P(\bm{ \sigma}) \simeq \int d\bm{W}^*  e^{-\frac{1}{2}\bm{W}^* \cdot \bm{W}^*  } \prod_{t=1}^T 
 \left\{\frac{e^{\beta \sigma_i (t) \frac{1}{\sqrt{N}} \sum_j W^*_j \sigma_j(t-1) }}
{2 \cosh \left[ \frac{\beta }{\sqrt{N}} \sum_j W^*_j \sigma_j(t-1) \right] } \; p(\{\sigma_j(t)\}_{j\neq i}) \right\},
\end{equation}
where the Gaussian measure accounts for 
our prior knowledge on the teacher couplings distribution.
Hence, for large $N$, we are effectively dealing with the statistical mechanics of a 
learning problem for a binary classifier neural network
(aka logistic regression), where the 'input' data $\sigma_j(t-1)$ are used to predict the 'outputs'
$\sigma_i(t)$; the input varibales are independent for different $t$, but have nontrivial 'spatial' correlations given by the 
matrix $\bm{C}$. 
The calculation of the free energy follows the steps of replica calculations for 
perceptron learning problems \cite{Nishimori_book,Engel_book,Opper_Kinzel_1996}.
Averages over $\sigma_j(t)$ factorize over time and can be expressed through Gaussian fields $h_a$ 
for each replicated coupling variable $W_a$, and fields
$
u=\frac{1}{\sqrt{N}} \sum_j W^*_j \sigma_j(t-1)
$ for the teacher.
Under the replica symmetry assumption, which is plausible to be correct for 
convex cost functions, the covariances are expressed by order parameters
\begin{eqnarray}
\left\langle u^2 \right\rangle & =& \frac{1}{N} \sum_{ij} W_i^*C_{ij} W_j^*=1, \\
\left\langle h_a u \right\rangle & =&  \frac{1}{N} \sum_{ij} W^a_i C_{ij} W^*_j \doteq R ,\\
\left\langle h_a^2 \right\rangle & =&  \frac{1}{N} \sum_{ij} W^a_i C_{ij} W^a_j \doteq q_0, \\
\left\langle h_a h_b \right\rangle & =&  \frac{1}{N} \sum_{ij} W^a_i C_{ij} W^b_j \doteq q\qquad a\neq b 
\end{eqnarray}
and the free energy
(\ref{eq:free_def}) is computed as
(appendix \ref{sec:AppA}):
\begin{equation}
\begin{split}
F=- Extr_{q,R,q_0}  \frac{1}{\nu} & \left\{ \frac{1}{2} \frac{q_0-R^2}{q-q_0}- \frac{1}{2} \log (q-q_0) - \frac{1}{2N} Tr \log C \right. \\
& \left. + \alpha  \sum_{\sigma_0} \int 
 \mathcal{D}t  \mathcal{D}y \frac{e^{\beta \sigma_0 (\sqrt{1-\frac{R^2}{q}} t + \frac{R}{\sqrt{q}} y ) }}
{2 \cosh [\beta (\sqrt{1-\frac{R^2}{q}} t + \frac{R}{\sqrt{q}}y )]} \right.  \\
&\left. \log \int \mathcal{D}z  e^{- \nu \mathcal{E}(\sigma_0,\sqrt{q_0 -q} z + \sqrt{q} y)}  \right\}.
\label{eq:free_nu}
\end{split}
\end{equation}
The limit 
$\nu \to \infty$ will occur with  $q_0 \to q$, since the different solutions $\bm{W}$ have to converge to the same minimum.  In this limit, keeping
the quantity $x \doteq (q_0 - q )\nu$ finite, we finally get 
\begin{equation}
\begin{split}
F= -  &Extr_{q,R,x,z}\left\{ \frac{q-R^2}{2x} + \alpha  \sum_{\sigma_0}  \int 
 \mathcal{D}t  \mathcal{D}y \frac{e^{\beta \sigma_0 (\sqrt{1-\frac{R^2}{q}} t + \frac{R}{\sqrt{q}} y ) }}
{2 \cosh [\beta (\sqrt{1-\frac{R^2}{q}} t + \frac{R}{\sqrt{q}}y )]} \right. \\
& \left. \qquad \qquad  \qquad \left[ -\frac{z^2}{2} -\mathcal{E}(\sigma_0,\sqrt{x} z + \sqrt{q} y) \right] \right\}.
\label{eq:free_2}
\end{split}
\end{equation}
Remarkably, the explicit dependence of $F$ on the correlation matrix (last term in the first line of equation \ref{eq:free_nu})
drops when taking the limit $\nu \to \infty$. Hence, the result we get for $F$ and for the order parameters extremizing $F$
is the same that we would get if the spins over which we are computing the expectations were independent and the matrix $C$ was not included in the calculation.
Still, the correlation matrix affects the error through the parameters $\rho$ and $Q$ defined in (\ref{eq:def_err}), which are found to be 
(appendix \ref{sec:AppA})
\begin{eqnarray}
&\rho  &=  R, \\
&Q&=  R^2 + (q-R^2) \frac{1}{N} Tr \overline{\bm{C}^{-1}},
\label{eq:def_rho_Q}
\end{eqnarray}
where $R$ and $q$ are the order parametrs extremizing the free energy (\ref{eq:free_2}).
Inserting the above equations in (\ref{eq:def_err}) we find the following result
for the error:
\begin{equation}
\varepsilon =1-2 R + q+ (q-R^2) \left(\frac{1}{N} Tr \overline{\bm{C}^{-1}} -1\right).
\label{eq:error_def_formula}
\end{equation}
The last term represents the effect of the correlations of the data on the error and vanishes
when $\bm{C}$ equals the unit matrix. This term can be shown to be positive and leads to an increase
in error. In section \ref{sec:results} we will give explicit results for the error of the 
EMF algorithm.

\section{Statistics of correlation matrices}
\label{sec:stat_C}
In this section we show how one can compute the stationary value of the negative integer moment of the spin correlations
\begin{equation}
C_{-1} \equiv \lim_{t\to\infty}\lim_{N\to\infty} \frac{1}{N} \mbox{Tr} \overline{\bm{C}^{-1}(t)},
\label{moment-1}
\end{equation}
necessary for the estimation error (\ref{eq:error_def_formula}). 
Here the bar denotes 
expectation with respect to independent random Gaussian couplings with zero mean and variance
$1/N$. 
Our analysis begins with the time evolution for the 
correlation matrix $\bm{C}(t)$ assuming zero magnetisations $m_j(t) =0$.
Following \cite{Mezard_Sakellariou_2011}, we can assume that in the limit of large $N$ the random variables $g_i$ and $g_j$, where
$g_i = \sum_k J_{ik} \sigma_k(t)$, are zero mean Gaussian random variables 
with $\langle g_i g_j\rangle =  \sum_{kl} J_{ik} C_{kl}(t) J_{lj}$ and $\langle g_i^2\rangle = 1$. An expansion with respect to weak correlations similar to equations (15-16) in \cite{Mezard_Sakellariou_2011}
yields the time evolution
\begin{equation}
\bm{C}(t+1) = \bm{I} \gamma(t) + a^2 \bm{J} \bm{C}(t) \bm{J^\top},
\end{equation}
where $\bm{I}$ is the unit matrix, $\bm{C}(0) = \bm{I}$ and $\bm{J} $ is the $N \times N$ coupling matrix.  The selfaveraging quantity $\gamma$ must be determined such that $C_{ii}(t) = 1$ yielding the condition that
$\gamma(t) = 1- a^2 \Tr \overline{ \bm{J} \bm{C}(t) \bm{J}^\top}$.   
Since we are interested in the stationary solution (25), we 
introduce the limiting value $\gamma\doteq \lim_{t\to\infty} \gamma(t) $ and
define  $\bm{B}(t) = \frac{1}{\gamma} \bm{C}(t)$, obtaining the
simplified iteration
\begin{equation}
\bm{B}(t+1) = \bm{I}  + a^2 \bm{J} \bm{B}(t) \bm{J^\top}, \qquad \mbox{having the solution}\qquad \bm{B}(t) = 
\sum_{k=0}^t a^{2k} \bm{J}^{k} (\bm{J^\top})^{k}.
\label{Corr_dynamics2}
\end{equation}
Note that in the limit of small $\beta$ (small $a$) one could choose to truncate the series in (\ref{Corr_dynamics2})
to the first order in $a$ (corresponding to k=0) and thus approximatig $\bm{B}$ by the unit matrix,
or to keep the first two orders in $a$ (up to $k=1$) and thus getting the sum of the unit matrix and a Wishart matrix.
From the above equations we get $\gamma = \frac{1}{1+a^2}$.  We can 
use (\ref{Corr_dynamics2}) to derive an iteration for the 
generating function of integer moments.
In the thermodynamic limit the calculation simplifies remarkably. Consider e.g. the
computation of $\lim_{N\to\infty}\frac{1}{N} \mbox{Tr} \overline{\bm{B}^k(t+1)}$ for some integer $k$. One would have to deal
with terms of the form
\begin{equation}
\frac{1}{N} \mbox{Tr} (\overline{\bm{J} \bm{B}(t) \bm{J^\top} \bm{J} \bm{B}(t) \bm{J^\top}  \cdots  \bm{J} \bm{B}(t) \bm{J^\top}}).
\label{graphs}
\end{equation}
Given the Gaussian form of the $\bm{J}$ random matrix, Wick's theorem applies and the expectation in (\ref{graphs}) can be computed using 
diagrammatic techniques. 
As is well known\cite{Hooft}, for $N\to\infty$ only the planar diagrams, i.e. the ones
for which lines are not crossing, will contribute to the limit. Besides,
note that in the evaluation of (\ref{graphs})
the terms containig 
$\contraction{}{\bm{J}}{\dotsc}{\bm{J}} \bm{J} \dotsc \bm{J} $ and
$\contraction{}{\bm{J}^\top}{\dotsc}{\bm{J}^\top} \bm{J}^\top \dotsc \bm{J}^\top $ 
pairings will vanish because of the asymmetry of the  $\bm{J}$ matrix. 
It is easy to see (an example is given in Appendix \ref{App_graph}) that 
this implies that also pairings of the kind 
$\contraction{}{\bm{B}(t)}{\dotsc}{\bm{J}} \bm{B}(t) \dotsc \bm{J} $ and 
$\contraction{}{\bm{B}(t)}{\dotsc}{\bm{J}^\top} \bm{B}(t) \dotsc \bm{J}^\top $ 
are forbidden.
Hence, in computing moments by iteration over time, we can treat $\bm{J}^{k}$ as
independent from $\bm{B}(t)$.
 We will not pursue the diagrammatic approach further 
but use this independence directly in the selfconsistent  computation of the generating function $S(x)$ of the 
asymptotic integer moments. This is given by
\begin{equation}
S(x) = \lim_{t\to\infty} S_{t}(x) = \sum_{k =0}^\infty (-x)^k  B_k , 
\label{def_genfu}
\end{equation}
where 
\begin{eqnarray}
\label{eq:mom_b}
S_{t}(x) \doteq \lim_{N\to\infty} \frac{1}{N} \mbox{Tr} \overline{\left(I + x \bm{B}(t)\right)^{-1}} , \\
B_k = \lim_{t\to\infty} \lim_{N\to\infty}\frac{1}{N} \mbox{Tr} \overline{\bm{B}^k(t)}.
\nonumber 
\end{eqnarray}
Finally, from $S(x)$ we can also deduce (\ref{moment-1})
\begin{equation}
C_{-1} = \frac{1}{\gamma} \lim_{x\to\infty} x S(x).
\end{equation}
We use an expression for $S_{t}(x)$ 
based on the Gaussian ensemble of auxiliary $N$- dimensional vectors $\bm{y}$. This is
 defined by the partition function
\begin{equation}
\begin{split}
Z_{t+1} (x) &= \int \prod_i dy_i \exp\left[-\frac{1}{2} \bm{y}^\top (I + x \bm{B}(t+1)) \bm{y}\right] = \\
&\int \prod_i dy_i \exp\left[-\frac{1}{2} (1 + x) \bm{y}^\top \bm{y} - \frac{a^2 x}{2} \bm{y}^\top \bm{J} \bm{B}(t) \bm{J}^\top \bm{y}\right],
\label{Gausspart_B}
\end{split}
\end{equation}
from which the generating function is obtained as
\begin{equation}
S_{t+1}(x) = \lim_{N\to\infty} \frac{1}{N}  \overline{\langle \bm{y}^\top \bm{y} \rangle}_{t+1},
\label{genfu_gauss}
\end{equation}
where the brackets denote expectation wrt to (\ref{Gausspart_B}). We 
compute the average over random matrices $\bm{J}$, using the fact that we can neglect the
dependency between the random matrices $\bm{J}$ and $\bm{B}(t)$ in the partition function (\ref{Gausspart_B}). 
An annealed average of (\ref{Gausspart_B}) and the limit $t\to\infty$
(appendix \ref{sec:AppB}) yields the self consistent equation
\begin{equation}
S(x) = \frac{1}{1+ x} S\left(a^2 x S(x)\right) .
\label{recurs_genfu}
\end{equation}
The explicit computation of moments is facilitated by introducing an auxiliary function $\phi$, its power series expansion (whose coefficients are denoted by $M_k$) and its inverse by
\begin{eqnarray}
\label{def_phi}
\phi(x) = \frac{a^2 x}{a^2 - x} S\left(\frac{x}{a^2 - x}\right) = x \sum_{k=0}^\infty (-1)^k x^{k} M_k,  \\
\label{def_phi_2}
a^2 y S(y) = \phi\left(\frac{a^2 y}{1 + y}\right).
\end{eqnarray}

\begin{figure}
 \vspace{8mm}
\centering
\includegraphics[width=0.57\textwidth]{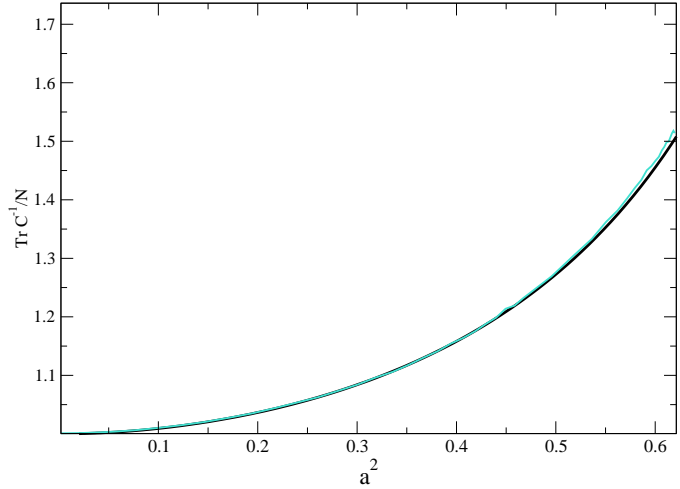}
\caption{The analytic result(black line) for $C_{-1} = \frac{1}{N} Tr \overline{C^{-1}}$ is compared
 with the values obtained from simulation (blue line) for $N= 200$ and $T= 1000$. 
Results are averaged over $50$ istances of the network and  error bars are negligible.}
\label{fig:Q}
\end{figure}

From (\ref{def_genfu}), (\ref{def_phi}) and taking the limit $y\to\infty$ in
(\ref{def_phi_2}), we obtain
\begin{equation}
C_{-1} = \frac{1}{\gamma a^2}\phi(a^2) = \frac{1}{\gamma}
\sum_{k=0}^\infty (-a^2)^k M_k .
\label{seriesB-1}
\end{equation}
We will next see how to obtain closed form expressions for the $B_k$ and $M_k$ recursively.
Let us first show that for known values of $B_1,\ldots, B_n$, we can compute $M_n$.
From (\ref{recurs_genfu}) and (\ref{def_phi}) we get the expression
\begin{equation}
\phi(x) = x S(\phi(x)).
\label{comp_phi}
\end{equation}
Applying Lagrange's inversion formula \cite{Wilf_book} to (\ref{comp_phi}) one can express the coefficients
of the power series expansion of $\phi(x)$ in terms of those of $S$:
\begin{equation}
M_n = \frac{(-1)^n}{n+1} [\phi^{n}] \{(S(\phi))^{n+1}\} = \frac{(-1)^n}{n+1} [\phi^{n}]
\left\{ \left(\sum_{k =0}^\infty (-1)^k \phi^k B_k\right)^n    \right \},
\label{eq:M_coef}
\end{equation}
where $[\phi^{n}]$ denotes the coefficient of $\phi^{n}$ in a power series expansion of the mathematical 
expression in the brackets $\{\ldots\}$. Finally, 
we insert in (\ref{eq:M_coef}) the expansion of $S$ (\ref{def_genfu}).  One can see that the coefficients are of the form
\begin{equation}
M_n = B_n + f_n(B_1,\ldots, B_{n-1}),
\label{MB_relation1}
\end{equation}
where the functions $f_n$ can be computed in closed form for any $n$ with a computer algebra programme such as 
Mathematica.  To obtain a relation for $B_n$, we expand both sides of
(\ref{def_phi}) into powers of $y$. Using elementary properties of binomial coefficients and comparing coefficients
 of $y^n$  yields
the second explicit relation
\begin{equation}
B_n = \sum_{l=0}^n a^{2l} {n\choose l} M_l = a^{2n} M_n +  \sum_{l=0}^{n-1} a^{2l} {n\choose l} M_l.
\label{MB_relation2}
\end{equation}
Hence, inserting (\ref{MB_relation1}) into (\ref{MB_relation2}), we obtain
\begin{equation}
B_n = \frac{1}{1- a^{2n}}\left(a^{2n} f_n(B_1,\ldots, B_{n-1}) +\sum_{l=0}^{n-1} a^{2l} {n\choose l} M_l\right).
\end{equation}
Unfortunately, the series (\ref{seriesB-1}) turns out to be an asymptotic one. Coefficients $M_n$ diverge 
for $n\to\infty$ and
one has to use a regularisation method such as the Borel summation or the Pad\`e approximation in order to extract
a useful result out of a finite number of coefficients. 
We have resorted to the latter method (appendix \ref{sec:AppC}).
Our results obtained in this way are in excellent agreement with simulations of the kinetic Ising model
for $N= 200$ and $T= 1000$. Figure \ref{fig:Q} shows that for small values of $a$, i.e. small $\beta$, 
the matrix $\bf{C} \approx I$. 
For increasing $\beta$ also $\bf{C}_{-1}$ increases but remains finite. Note, that
for $\beta\to\infty$, the parameter $a$ converges to the value $a= \sqrt{2/ \pi}$.

\section{Results}
\label{sec:results}
\begin{figure}
 \vspace{8mm}
\centering
\includegraphics[width=0.57\textwidth]{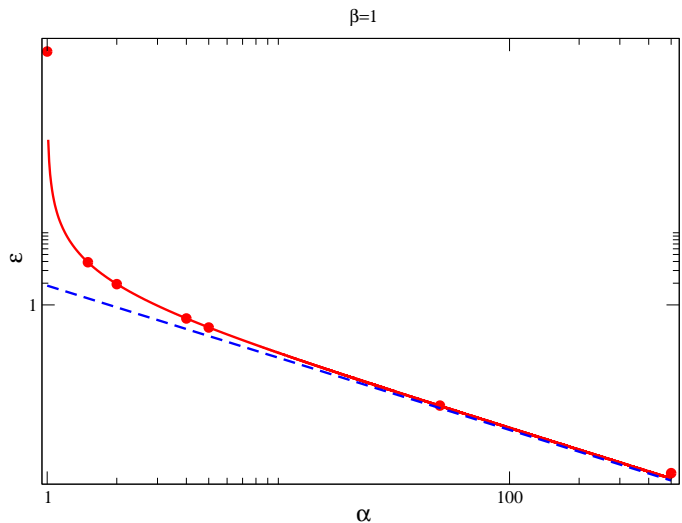}
\caption{Mean squared error of the couplings inferred with the EMF method (red dots)
for a system of size N=200 with $\beta=1$. Results are averaged over 25 istances of the network. Error bars are negligible. 
The red line corresponds to the replica
result for the EMF prediction error, the blue line to the replica result for the 
asymptotic optimal prediction error.}
\label{fig:Q_1}
\end{figure}

\begin{figure}
 \vspace{8mm}
\centering
\includegraphics[width=0.57\textwidth]{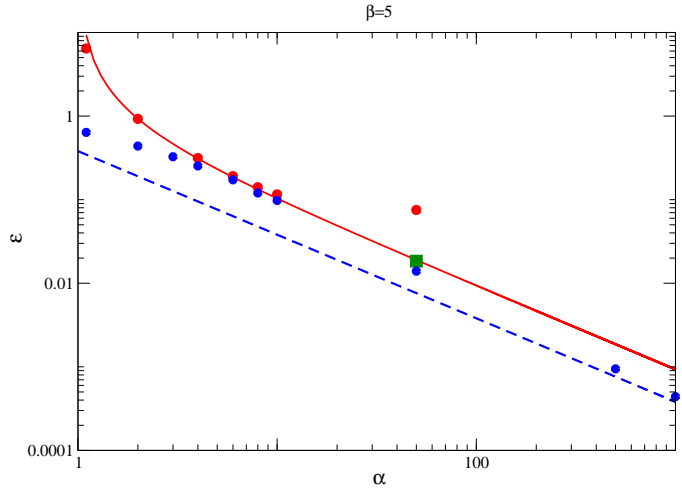}
\caption{Mean squared error of the couplings inferred with the EMF method (red dots)
for a system of size N=200 with $\beta=5$. Results are averaged over 25 instances of the network. 
The red line corresponds to the replica
result for the EMF prediction error, the blue line to the replica result for the 
optimal prediction error. The blue dots are results from simulations of a penalised ML algorithm. Error bars are negligible.
For large values of $\alpha$, the EMF method displays finite-size effects (see the red dot at $\alpha=50$), which 
are stronger for larger $\beta$. The green dot takes into accout finite-size corrections, and it is obtained as explained in figure \ref{fig:Q_3}.}
\label{fig:Q_2}
\end{figure}

In the case of the EMF estimator (\ref{eq:E_EMF}) the free energy (\ref{eq:free_2}) becomes
\begin{equation}
F= Extr_{R,q} \left\{ \frac{q-R^2}{2x} - \frac{\alpha}{1+ 2 a^2 x} \left( \frac{1}{2} +\frac{a^2q}{2}   
 - aR\int \mathcal{D}x  \: x \tanh (\beta x) \right)  \right\},
\end{equation}
and the extremum conditions yield the following equations for the order parameters:
\begin{align}
&R= 1 \\
&q= \frac{ a^2 (\alpha -2)+1}{a^2 (\alpha -1)} \\
&x= \frac{1}{2a^2 (\alpha -1)}.
\end{align}
Inserting the above equations in (\ref{eq:error_def_formula}) the error is computed as follows:
\begin{equation}
\varepsilon_{EMF} =\frac{1}{\alpha -1} \frac{ 1- a^2 }{a^2 }  \frac{1}{N} Tr \overline{C^{-1}}.
\label{prederr1}
\end{equation}
We defer a detailed analysis of the finite $\alpha$ performance of the ML estimator to
a future publication. Here we are interested in the leading behaviour of the decay of the 
prediction error as $\alpha\to\infty$. It is well known that ML estimators are asymptotically efficient, i.e.
the errors decay at an optimal speed. Hence, our asymptotic result should be a yardstick that allows
for a comparison of algorithms. 
The calculation in Appendix \ref{sec:AppE} shows that 
for large values of the  $\alpha$ parameter this optimal error decays as
\begin{equation}
\epsilon_{opt} \simeq  \frac{1}{\beta a \alpha} \frac{1}{N} Tr C^{-1}.
\label{prederr2}
\end{equation}
Hence, for $\alpha\to\infty$, we have 
\begin{equation}
\lim_{\alpha\to\infty} \frac{\epsilon_{opt}}{\epsilon_{EMF}} =   \frac{a}{\beta (1-a^2)}.
\label{ratio_error}
\end{equation}
For small $\beta$, i.e. large stochasticity of the spins, we have $a\simeq \beta$ and 
both algorithms decay at the same rate. This can still be seen in figure \ref{fig:Q_1} for $\beta=1$, where
the EMF algorithms performs close to optimal. For larger $\beta$, the spins behave more
deterministically and as shown in figure \ref{fig:Q_2}  the EMF algorithm deviates significantly from optimality.
We have also included data points from a  simulation of a penalised ML estimator, where we have
minimised the cost function $E_{ML} + \frac{\bm{W}^\top \bm{W}}{2}$ numerically by a gradient descent
algorithm. Note that the penalty term we chose is equivalent to the prior and we are thus maximizing the log-posteror.
One can see that this type of algorithm achieves asymptotic optimality. Finally,
with increasing $\beta$ the ratio (\ref{ratio_error}) decays to zero. While the decay rate of the EMF algorithm
converges to a nonzero value (note that for $\beta\to\infty$, we have $a\to \sqrt{2/ \pi}$), the optimal 
asymptotic error rate converges to zero indicating a transition to a faster decay than $1/\alpha$ in the limit.
It is also interesting to note that for larger $\beta$ simulations of the EMF algorithms show strong finite size 
effects in $N$ and the error reaches a plateau for increasing $\alpha$. Hence, we had to apply
a finite scaling for the last simulation point in figure \ref{fig:Q_2} .

\begin{figure}
 \vspace{8mm}
\centering
\includegraphics[width=0.57\textwidth]{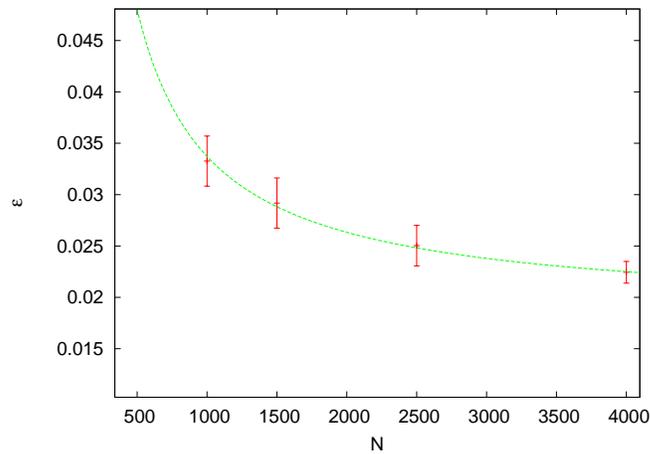}
\caption{EMF prediction error for fixed $\alpha=50$ and $\beta=5$  as a function of $N$. 
Fitting a power law to the data we find the asymptotic value valid for large $N$, which
 corresponds to the green dot in figure \ref{fig:Q_2}.}
\label{fig:Q_3}
\end{figure}

\section{Outlook}
It will be interesting to develop and study algorithms which include prior knowledge about
the couplings to be learnt. This could be done within a Bayesian approach where a prior probability
density over couplings is specified. In this way one may e.g. introduce sparsity. Using a similar replica
approach, one could compare the performance of different algorithms to that of the Bayes estimator, 
which is optimal on average over teacher networks drawn at random from the prior. A nontrivial question
is that of an algorithmic realisation of the Bayes predictor. We expect that cavity approaches 
(TAP equations) could be applied to get a tractable approximation which becomes exact in the thermodynamic limit. 
We also expect that one should include explicit knowledge of the statistics of the spin correlations into such an approach
in order to get optimal performance.

\section*{Acknowledgements}
This work is supported by the Marie Curie Training Network NETADIS (FP7, grant 290038).

\section*{Appendix}
\begin{appendices}

\section{Details of the replica calculation of the free energy}
\label{sec:AppA}
After some standard manipulations \cite{Nishimori_book, Engel_book, Opper_Kinzel_1996},
the quenched free energy (\ref{eq:free_def}) is computed as
\begin{equation}
\begin{split}
F=- Extr_{q,R,q_0}  \frac{1}{\nu} & \left\{G(R,q,q_0) + \alpha  \sum_{\sigma_0} \int 
 \mathcal{D}t  \mathcal{D}y \frac{e^{\beta \sigma_0 (\sqrt{1-\frac{R^2}{q}} t + \frac{R}{\sqrt{q}} y ) }}
{2 \cosh [\beta (\sqrt{1-\frac{R^2}{q}} t + \frac{R}{\sqrt{q}}y )]} \right.  \\
&\left. \log \int \mathcal{D}z  e^{- \nu \mathcal{E}(\sigma_0,\sqrt{q_0 -q} z + \sqrt{q} y)}  \right\},
\end{split}
\end{equation}
where $G(R,q,q_0)$ is the weight of the coupling vectors $\bm{W}$ which are constrained by the order parameters:
\begin{equation}
G(R,q,q_0) = \lim_{n \to 0} \frac{\partial}{\partial n} \frac{1}{N}\ln Z_{coup},
\end{equation}
with
\begin{equation}
\begin{split}
Z_{coup} &= \int d\bm{W}^* \prod_a d\bm{W^a}
e^{-\frac{1}{2} \bm{W}^* \cdot \bm{W}^*  }
 \prod_{a } \delta(\sum_{ij} W^a_i C_{ij} W^*_j - N q_0) \\
& \prod_{a } \delta(\sum_{ij} W^a_i C_{ij} W^a_j - N R)
 \prod_{a < b} \delta(\sum_{ij} W^a_i C_{ij} W^b_j - Nq). \\
\end{split}
\end{equation}
We can decouple the integrals over different spins  by diagonalising 
$C = U\Lambda U^\top$ and transforming to new variables 
$U^\top W^a \to W^a$, 
$U^\top W^* \to W^*$
which we give just the same name:
\begin{equation}
\begin{split}
Z_{coup} &= \int d\bm{W}^* \prod_a d\bm{W^a}
e^{-\frac{1}{2} \bm{W}^* \cdot \bm{W}^*  }
 \prod_{a } \delta(\sum_{i} W^a_i  \Lambda_i W^*_i  - N q_0) \\
& \prod_{a } \delta(\sum_{i} W^a_i \Lambda_i W^a_i - N R)
 \prod_{a < b} \delta(\sum_{i} W^a_i \Lambda_i W^b_i - Nq). \\
\label{eq:Z_coup}
\end{split}
\end{equation}
The integration over the couplings and the auxiliary parameters gives rise to the following equation for $G$:
\begin{equation}
G(R,q,q_0) = \frac{1}{2} \frac{q_0-R^2}{q-q_0}- \frac{1}{2} \log (q-q_0) - \frac{1}{2N} Tr \log C.
\end{equation}
In order to compute the parameters $\rho$ and $Q$ 
 from the free energy $F$ , we introduce the auxiliary variables $\{ \eta_1, \eta_2 \}$ in 
the partition function $Z_{coup}$ 
(\ref{eq:Z_coup}) as follows:
\begin{equation}
\begin{split}
Z_{coup} = &\int d\bm{W}^* \prod_a d\bm{W^a}\; d \hat{q_0} \; d\hat{R} \; d\hat{q} \; 
e^{-\frac{1}{2}\bm{W}^* \cdot \bm{W}^*  }
 \prod_{a } e^{i\hat{q_0}(\sum_{i} W^a_i  \Lambda_i W^*_i - N q_0)} \\
& \prod_{a } e^{i\hat{R}(\sum_{i} W^a_i  (\Lambda_i+\eta_1) W^a_i - N R)}
 \prod_{a < b} e^{i\hat{q}(\sum_{i} W^a_i (\Lambda_i+\eta_2) W^b_i - Nq) }.
\end{split}
\end{equation}
By derivatives with respect to $\{ \eta_1, \eta_2 \}$
and taking the limit $\eta_1 \to 0, \eta_2 \to 0$ one recovers (\ref{eq:def_rho_Q}).

\section{Derivation  of the generating function}
\label{sec:AppB}
For a Gaussian model without external field we have 
$\langle {y_i} \rangle = 0$,
hence
$ q= \frac{1}{N} \sum_i \langle {y_i} \rangle^2 = 0$
and there is no need to introduce replicas,
(absence of spin--glass ordering) and we can restrict ourselves to an annealed average.
Decoupling the quadratic form in the exponent of (\ref{Gausspart_B}) 
using correlated Gaussian random vectors with covariance $\langle \bm{z} \bm{z}^\top\rangle_c = \bm{B}(t)$,
we get
\begin{eqnarray}
\nonumber
\overline{Z_{t+1} (x)}  = \int \prod_i dy_i \exp\left[-\frac{1}{2} (1 + x) \bm{y}^\top \bm{y}\right] 
\left\langle \exp\left( - \frac{a^2 x}{2N} (\bm{z}^\top \bm{z}) (\bm{y}^\top \bm{y}) 
\right)\right\rangle_z \\
\propto \int_0^\infty ds\; s^{\frac{N+1}{2}} \exp[ -\frac{N}{2}(1+x) s]\;  \left\langle\exp( - \frac{a^2 x}{2N} (\bm{z}^\top \bm{z}) s \right\rangle_z \\
\nonumber
\propto \int_0^\infty ds\; s^{\frac{N+1}{2}} \exp[ -\frac{N}{2}(1+x) s]\;  \left| I + a^2 x s \bm{B}(t)\right|^{-1/2} \\
\nonumber
=  \int_0^\infty ds\; s^{\frac{N+1}{2}} \exp\left[ -\frac{N}{2}(1+x) s - \frac{1}{2} \mbox{Tr} \ln( I + a^2 x s \bm{B}(t))\right],
\end{eqnarray}
where in the second line we have introduced polar coordinates $s= \frac{1}{N} \bm{y}^\top \bm{y}$. We compute the final integral for $N\to\infty$
by Laplace's method, and use the fact that from (\ref{genfu_gauss}) the maximiser of the integral gives
$s = \frac{1}{N} \langle \bm{y}^\top \bm{y}\rangle = S_{t+1}(x)$.
Finally from
$- \frac{1}{2} \mbox{Tr} \ln( I + a^2 x s \bm{B}(t)) = \mbox{const} + \ln Z_t(a^2 x s)$ we get the recursion
\begin{equation}
S_{t+1}(x) = \frac{1}{1+ x} S_t\left(a^2 x S_{t+1}(x)\right) .
\end{equation}
Taking the limit $t\to\infty$ yields (\ref{recurs_genfu}).

\section{Independence of the $\bm{J}$ and $\bm{B}(t)$ matrices: an example}
\label{App_graph}
To better illustrate the independence of the $\bm{J}$ and $\bm{B}(t)$ matrices,
let us give an example and consider the evaluation of one of the terms needed for the computation of 
$\lim_{N\to\infty}\frac{1}{N} \mbox{Tr} \overline{\bm{B}^k(t+1)}$ 
(see \ref{graphs}):
\begin{equation}
\frac{1}{N} \mbox{Tr} (\overline{\bm{J} \bm{B}(t) \bm{J^\top}  \bm{J} \bm{B}(t) \bm{J^\top}}).
\end{equation}
The only sets of contractions giving nonzero contribution in the large $N$ limit are the following two:
\begin{equation}
\begin{split}
&\contraction[2ex]{\frac{1}{N} \mbox{Tr} (}{\bm{J}}{\bm{B}(t) }{\bm{J^\top}}
\contraction[2ex]{\frac{1}{N} \mbox{Tr} ( \bm{J} \bm{B}(t) \bm{J^\top}  }{\bm{J}}{\bm{B}(t)}{\bm{J^\top})}
\frac{1}{N} \mbox{Tr} ( \bm{J} \overline{\bm{B}}(t) \bm{J^\top}  \bm{J} \overline{\bm{B}}(t)  \bm{J^\top})=\frac{1}{N}\mbox{Tr} \left(\overline{\bm{B}}(t)\right)^2 ,\\
\contraction{\frac{1}{N} \mbox{Tr} ( \bm{J} \bm{B}(t)}{\bm{J}}{}{\bm{B}(t)}
&\contraction[2ex]{\frac{1}{N} \mbox{Tr} (\bm{J} }{\bm{B}(t)}{ \bm{J^\top}  \bm{J} }{\bm{B}(t)}
\contraction[3ex]{\frac{1}{N} \mbox{Tr} ( }{\bm{J}}{\bm{B}(t) \bm{J^\top}  \bm{J} \bm{B}(t)}{\bm{J^\top})}
\frac{1}{N} \mbox{Tr} ( \bm{J} \bm{B}(t) \bm{J^\top}  \bm{J} \bm{B}(t) \bm{J^\top})=\frac{1}{N}\mbox{Tr} \overline{(\bm{B}(t)^2 )}.
\end{split}
\end{equation}
The contractions involving the pairing of a $\bm{J}$ with a  $\bm{B}(t) $
vanish, since they involve either 
$\contraction{}{\bm{J}}{\dotsc}{\bm{J}} \bm{J} \dotsc \bm{J} $ 
($\contraction{}{\bm{J}^\top}{\dotsc}{\bm{J}^\top} \bm{J}^\top \dotsc \bm{J}^\top $ )
pairings or crossing lines (resulting in non planar diagrams),
as shown in the two examples below:
\begin{equation}
\contraction{\frac{1}{N} \mbox{Tr} ( \bm{J} }{\bm{B}(t)}{}{\bm{J^\top}}
\contraction{\frac{1}{N} \mbox{Tr} ( \bm{J} \bm{B}(t) \bm{J^\top} }{\bm{B}(t)}{}{\bm{J^\top})}
\contraction[2ex]{\frac{1}{N} \mbox{Tr} (}{ \bm{J}}{\bm{B}(t) \bm{J^\top}}{\bm{J}}
\frac{1}{N} \mbox{Tr} ( \bm{J} \bm{B}(t) \bm{J^\top}  \bm{J} \bm{B}(t) \bm{J^\top})=0,
\qquad
\contraction{\frac{1}{N} \mbox{Tr} ( \bm{J} }{\bm{B}(t)}{}{\bm{J^\top}}
\contraction{\frac{1}{N} \mbox{Tr} ( \bm{J} \bm{B}(t) \bm{J^\top}  }{\bm{J}}{\bm{B}(t)}{\bm{J^\top})}
\contraction[2ex]{\frac{1}{N} \mbox{Tr} (}{\bm{J}}{\bm{B}(t) \bm{J^\top}  \bm{J} }{\bm{B}(t) }
\frac{1}{N} \mbox{Tr} ( \bm{J} \bm{B}(t) \bm{J^\top}  \bm{J} \bm{B}(t) \bm{J^\top})=0.
\qquad
\end{equation}

\section{Pad\`e Approximant}
\label{sec:AppC}

The so called Pad\`e approximant \cite{Numerical_recipes}, is a rational function (of a specified order)
whose power series expansion agrees with a given power series to the highest possible order.
Given a rational function of the form
\begin{equation}
R(x) \equiv \sum_{k=0}^M a_k x^k \left/ \left(  1+ \sum_{k=1}^N b_k x^k \right) \right.,
\end{equation}
then $R$ is said to be the Pad\`e approximant to the series
\begin{equation}
f(x) = \sum_{k=0}^{\infty} c_k x^k 
\end{equation}
if the following set equations is satisfied:
\begin{eqnarray}
&R(0)=f(0) \\
&\left. \frac{d^k}{d \,x^k}R(x) \right\rvert_{x=0}   = \left. \frac{d^k}{d \,x^k} f(x) \right\rvert_{x=0}  \quad k=1, \dots,M+N ,
\end{eqnarray}
which gives $M+N+1$ equations for the unknowns $a_0, \dots, a_M$ and $b_0, \dots, b_N$.

\section{Details on the statistics of the correlation matrix}
\label{sec:AppD}
The iterative methods explained is section \ref{sec:stat_C} allows us to calculate the moments $B_k$
and $M_k$, defined respectively in (\ref{eq:mom_b}) and (\ref{def_phi}),
for any given $k$. As an example, in the following we will enumerate the first three moments.
 \begin{eqnarray}
B_1&=&(1-a^2)^{-1} \\
B_2&=&(1-a^4)^{-1}(1-a^2)^{-2} \\
B_3&=&(1+2a^4)(1-a^6)^{-1}(1-a^4)^{-1}(1-a^2)^{-3} \\
M_1&=&(1-a^2)^{-1} \\
M_2&=&(2-a^4) (1-a^2)^{-2}(1-a^4)^{-1} \\
M_3&=& (5+a^4-4a^6+a^{10})(1-a^2)^{-4}(1-a^4)^{-1}(1+a^2+a^4).
\end{eqnarray}

\section{Asymptotic order parameters for ML estimator}
\label{sec:AppE}
The free energy for the ML estimator is given by
\begin{equation}
\begin{split}
F= - &Extr_{q,R,x, z} \left\{ \frac{q-R^2}{2x} + \alpha  \sum_{\sigma} \int 
 \mathcal{D}t  \mathcal{D}y \frac{e^{\beta \sigma (\sqrt{1-\frac{R^2}{q}} t + \frac{R}{\sqrt{q}} y ) }}
{2 \cosh [\beta (\sqrt{1-\frac{R^2}{q}} t + \frac{R}{\sqrt{q}}y )]}  \right.\\
& \left. \left[-\frac{z^2}{2} + \beta \sigma (\sqrt{x} z + \sqrt{q} y) -\log 2 \cosh [\beta (\sqrt{x} z + \sqrt{q} y)] \right] \right\} .
\end{split}
\label{free:ML1}
\end{equation}
It is possible to show that for $\alpha\to\infty$ one can assume that $q- R^2\to 0$, $x\to 0$ and $q\to 1$.
 Expanding the $\alpha$
dependent part of (\ref{free:ML1}) for small $\sqrt{x}$, solving for $z$ and finally taking the limit $q\to R^2$, we obtain
\begin{equation}
F \simeq - Extr_{q,R,x} \left\{ \frac{q-R^2}{2x} + \alpha 
\left(\frac{\beta a x}{2} + R b + \int \mathcal{D}y \log 2 \cosh [\beta\sqrt{q} y)] \right)\right\} .
\end{equation}
 This yields the following asymptotic scaling of
order parameters:
\begin{equation}
R\simeq 1,\quad x\simeq \frac{1}{\alpha b}, \quad q-R^2 \simeq \frac{1}{\alpha b}.
\end{equation}
Inserting the above expressions in the definition (\ref{eq:error_def_formula}) one obtains
 (\ref{prederr2}).

\end{appendices}
\section*{References}
\bibliographystyle{fbs}       
\bibliography{References.bib}   


\end{document}